\documentclass[runningheads]{llncs}
\usepackage{graphicx}
\usepackage{todonotes}
\usepackage{caption}
\usepackage{subcaption}
\usepackage[T1]{fontenc}

\begin{document}
\title{Deploying Enhanced Speech Feature Decreased Audio Complaints at SVT Play VOD Service\thanks{Supported by The Swedish Association of Hard of Hearing People.}}
\titlerunning{Enhanced Speech Feature Decreased Audio Complaints at SVT Play}
%
\author{Annika Bidner\inst{1}\orcidID{0000-0001-5878-1020}\and
Julia Lindberg\inst{1}\orcidID{0000-0001-8028-9788}\and\\
Olof Lindman\inst{1}\orcidID{0000-0001-9800-9318}\and
Kinga Skorupska\inst{2}\orcidID{0000-0002-9005-0348}}
\authorrunning{A. Bidner et al.}
%
\institute{SVT, Stockholm 105 10, Sweden
\email{annika.bidner@svt.se}\\
\url{https://www.svtplay.se/genre/tydligare-tal}
\and
Polish-Japanese Academy of Information Technology, XR Lab
}
\maketitle              
\begin{abstract}
At Public Service Broadcaster SVT in Sweden, background music and sounds in programs have for many years been one of the most common complaints from the viewers. The most sensitive group are people with hearing disabilities, but many others also find background sounds annoying. To address this problem SVT has added Enhanced Speech, a feature with lower background noise, to a number of TV programs in VOD service SVT Play. As a result, when the number of programs with the Enhanced Speech feature increased, the level of audio complaints to customer service decreased. The Enhanced Speech feature got the rating 8.3/10 in a survey with 86 participants. The rating for possible future usage was 9.0/10. In this article we describe this feature's design and development process, as well as its technical specification, limitations and future development opportunities.


\end{abstract}
\section{Introduction and Related Works}

SVT is the Public Service TV Broadcaster in Sweden. According to SVT's broadcasting licence \cite{broadcastinglicence} and audibility policy from 2020, SVT should prioritise good audibility by taking into account that background sound can make it harder for people with hearing loss to hear the program dialogue. The audibility work is fundamental for the public service offer to be accessible to everyone in Sweden. In 2020, about 1.5 million people in Sweden experienced hearing loss, 18 percent of the population over 16 \cite{hearingloss}. A similar commitment is seen in the European Accessibility Act.

People with hearing disabilities are sensitive to background noise, but many others also find background sounds annoying, for example when loud sounds could disturb neighbours, or when the viewing environment is noisy. Shirley et al \cite{shirley_intelligibility_2020} discuss some of the problems with the experience of speech in broadcast audio and many of these points are also applicable to TV broadcasts and streaming. There are several other techniques to address the problem of audio accessibility. All programs at SVT get subtitles, except short clips and live broadcasts. But this won't help if the viewer has dyslexia or is visually impaired. Dubbing is used for foreign content in many countries, but this is mainly done in kids' shows in Sweden. A study from 2012 showed that most participants strongly dislike dubbing, and tend not to use it. \cite{fors_2012}. At SVT, background music and sounds in programs have for many years been the most common type of complaint from the viewers. These two quoted complaints represent common problems:

\textbf{Complaint 1:} \textit{"Hello! I'm listening to the program on the Swedish military plane efforts in Afghanistan. I've written to you before about these constant background noises in your programs. This program is the worst so far, considering the annoying airplane noise and the "excuse me" goddamn piano tinkle. I don't know who handles my message but I know that someone responsible should listen to the program and try to realise how people with hearing loss experience this background sound. We have the right to perceive what is said in a program. I might contact The Swedish Association of Hard of Hearing People to get some improvements in this area."}

\textbf{Complaint 2:} \textit{"WHY must the background music be so loud, AT THE SAME TIME as people are talking in a program???? My hearing is normal, but it's Impossible to hear what they are saying. Luckily, there are subtitles, so i can use that to understand what is said. The background noise is sooo annoying and distracting. It takes away the pleasure of viewing the show I want to see. Is the noise really necessary?"}

In 2019, SVT received more than a hundred sound complaints per week. SVT often adjust the sound mix of programs that get complaints, but this hasn't reduced the number of complaints. Even so, no real progress had been made in SVT Play, SVT's VOD service, to improve audibility of speech in TV programs, until the experimental project, described in this paper, started in January 2020. 

\textbf{The main hypothesis underlying the project was: If SVT can provide an audio feature that makes it easier to hear the dialogue, and include it in a substantial number of programs, many people will use it and the complaints to SVT on audibility will decrease.
} 

An obvious alternative to enhancing the speech in videos would be to keep background sounds in programs to a minimum. But most viewers enjoy music and relevant sounds in programs. It helps create a more emotional and immersive experience \cite{music}. Thus, BBC Research and Development work on enhanced audio tested on "Casualty" drama series \cite{ward2019casualty} became an inspiration for the project. BBC tested an Accessible and Enhanced Audio Mix in the form of a slider on the Taster platform, in 2019 \cite{bbcnews_2019}, but so far, it has not been released as a standard feature. Another source of inspiration was the German Fraunhofer institute, that produces technical solutions for enhancing the human voice and reducing background noise \cite{Frauiis_2020}. 

In the first phase of the project, SVT representatives contacted different stakeholders, and discussed current audibility research and possible technical directions to enhance speech in TV programs. First, ORCA Europe, a Hearing Research Laboratory in Stockholm. SVT considered strengthening the main speech frequencies to make the dialogue more prominent. Florian Wolters from ORCA Europe discouraged SVT from doing that: "Raising certain frequencies in the dialogue will make the sound unnatural. It should be as natural as possible." \cite{signaltonoise_auto} Josefina Larsson, also from ORCA, said: "To lower the background sounds is a good idea. That should include all sounds that isn't notional dialogue." Alf Lindberg from SAHHP said "Avoid sudden changes in volume. Hearing aids adapt to a certain sound level, and users don't get a good experience when background noise is raised between dialogue passages." He promoted a continuous audio mix, to cater to different degrees of hearing loss. SVT has also been in contact with Dolby, a US company specializing in audio noise reduction and encoding. \cite{dolbyioblogrss} Dolby's solution "Dialog enhancement" can be applied to stereo sound, but it could not separate speech in the way SVT was aiming for.  

\section{Methods}

The first approach in the project was to manipulate the sound of video content by enhancing the main sound frequency bands (1- 4kHz) for human speech. A prototype was made and tested within the project group, but it was clear that this method did not fulfill its purpose. Background sounds within the voice frequency band were also enhanced and some important parts of speech with frequencies outside the span were lost. According to hearing experts at ORCA Europe, this solution was questionable, as it would distort the sound, so this approach was abandoned. The second approach was to use video with multichannel audio (5.1 or 3.0 sound) and adjust the balance between the different channels in the downmixing process, with notional speech in the center channel. The recommendation from hearing experts was to enhance the speech relative to the background by decreasing the volume of background sounds. New prototypes were created, with either -3 or -10 dB background noise. 

\begin{figure}[t!]
\includegraphics[width=\textwidth]{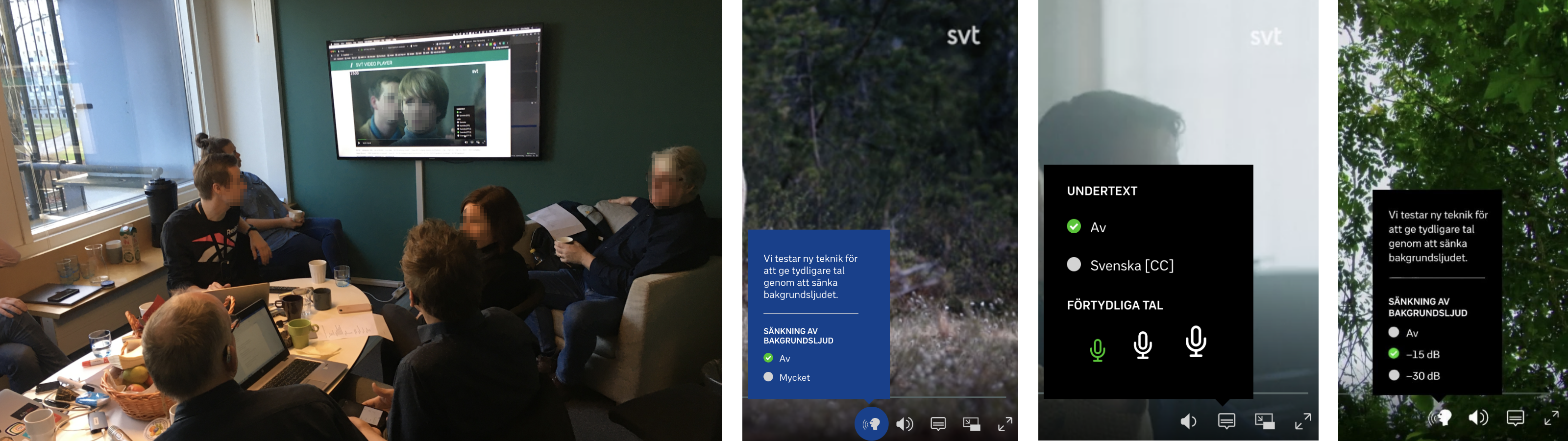}
\caption{Left to Right: Testing and iteration of the first prototype, January 10, 2020; Prototype 1: on/off; Prototype 2: Three steps; Prototype 3: Demo version with -15 and -30 dB} \label{testing}
\end{figure}

These prototypes were used in a participatory design session with representatives from the The Swedish Association of Hard of Hearing People along with developers, sound technicians, UX designers and accessibility specialists. It was clear that the reduction of background sound was not sufficient. The quality of the sound separation in channels as well as legal rights for manipulating sound in video content were issues that needed to be solved.

The term Enhanced Speech (Tydligare tal) was selected in cooperation with Tillgänglighetstruppen, an accessibility Facebook group with 2,000 members. 

They also took part in user testing. The prototype UI was iterated after feedback from users. The two sound alternatives -15 dB -30 dB, were considered too complicated and technical, so those were merged into one. This had technical benefits, since only one extra audio track had to be produced for each episode. The result was this UI, used in production in svtplay.se.

\begin{figure}[t!]
\includegraphics[width=\textwidth]{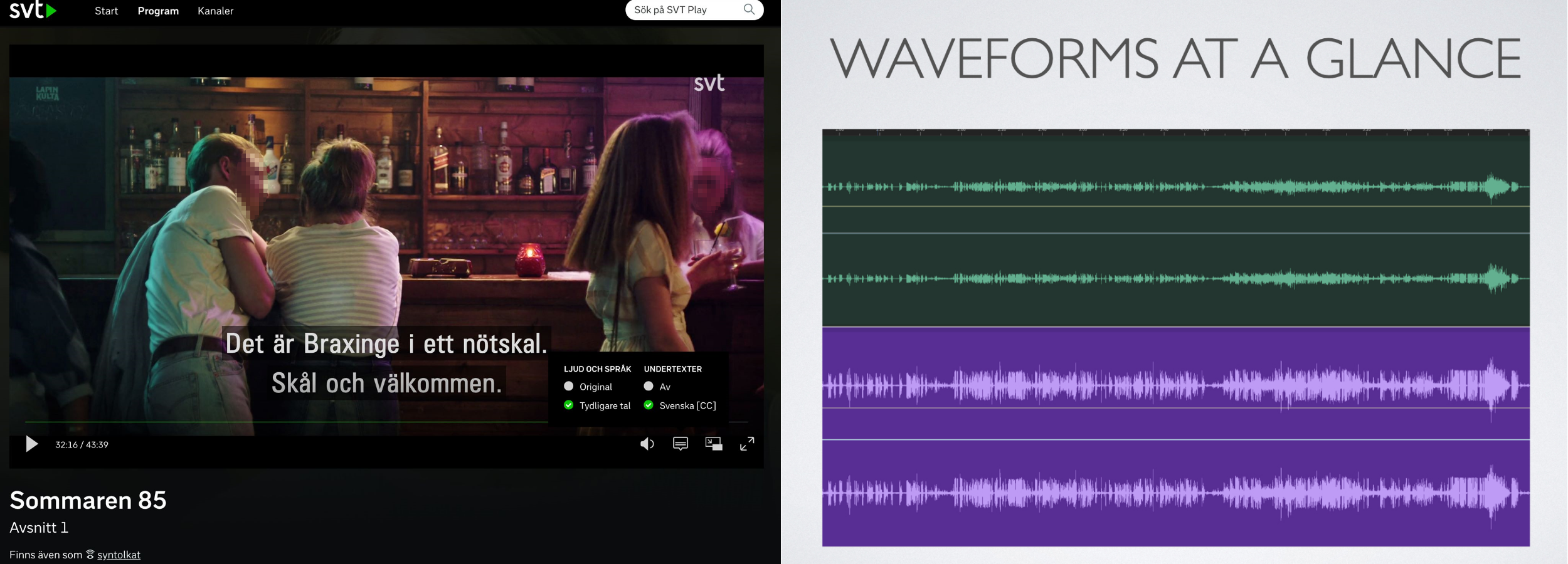}
\caption{Left: UI used in SVT Play, September 2020; Right: Stereo waveform of regular downmix (Upper green) and Enhanced Speech downmix (Lower purple)} \label{testing}
\end{figure}

\subsection{Technical Description}
On a technical level, Enhanced Speech is achieved by utilizing an audio panning filter, to create a downmix matrix that converts predefined multichannel audio into a dialogue enhanced audio track. Specifically, the desired audio output is accomplished by modifying the digital gain of each individual audio channel in a remixing or downmixing process. The parameters used to decide the amount of gain is defined as decimal multiples rather than a dB values, and in effect the procedure alters the sound strength of each individual channel prior to the actual mixing. Before the modified input channels are merged together a programmatic check is made to measure if either of the resulting tracks are going to clip above 0 dBFS, and if that is the case the whole track will be normalized by max peak when mixed.
The configuration of the panning filter is static, which means that a single downmix matrix is used for all content and that the parameters are always the same. For multichannel to stereo conversions the matrix is as follows:

\[L_{ES} <= (0.25* L) + (1.5 * C) + (0.25 * Ls)\]
\[R_{ES} <= (0.25* R) + (1.5 * C) + (0.25 * Rs)\]

\noindent Whereas a normal downmix from 5.1 surround to 2.0 stereo, as defined in the EBU R 128 recommendation (https://tech.ebu.ch/docs/r/r128.pdf), would use the following values:

\[L_{Stereo} = (1* L) + (0.707 * C) + (0.707 * Ls)\]
\[R_{Stereo} = (1* R) + (0.707 * C) + (0.707 * Rs)\]

\begin{figure}
\includegraphics[width=\textwidth]{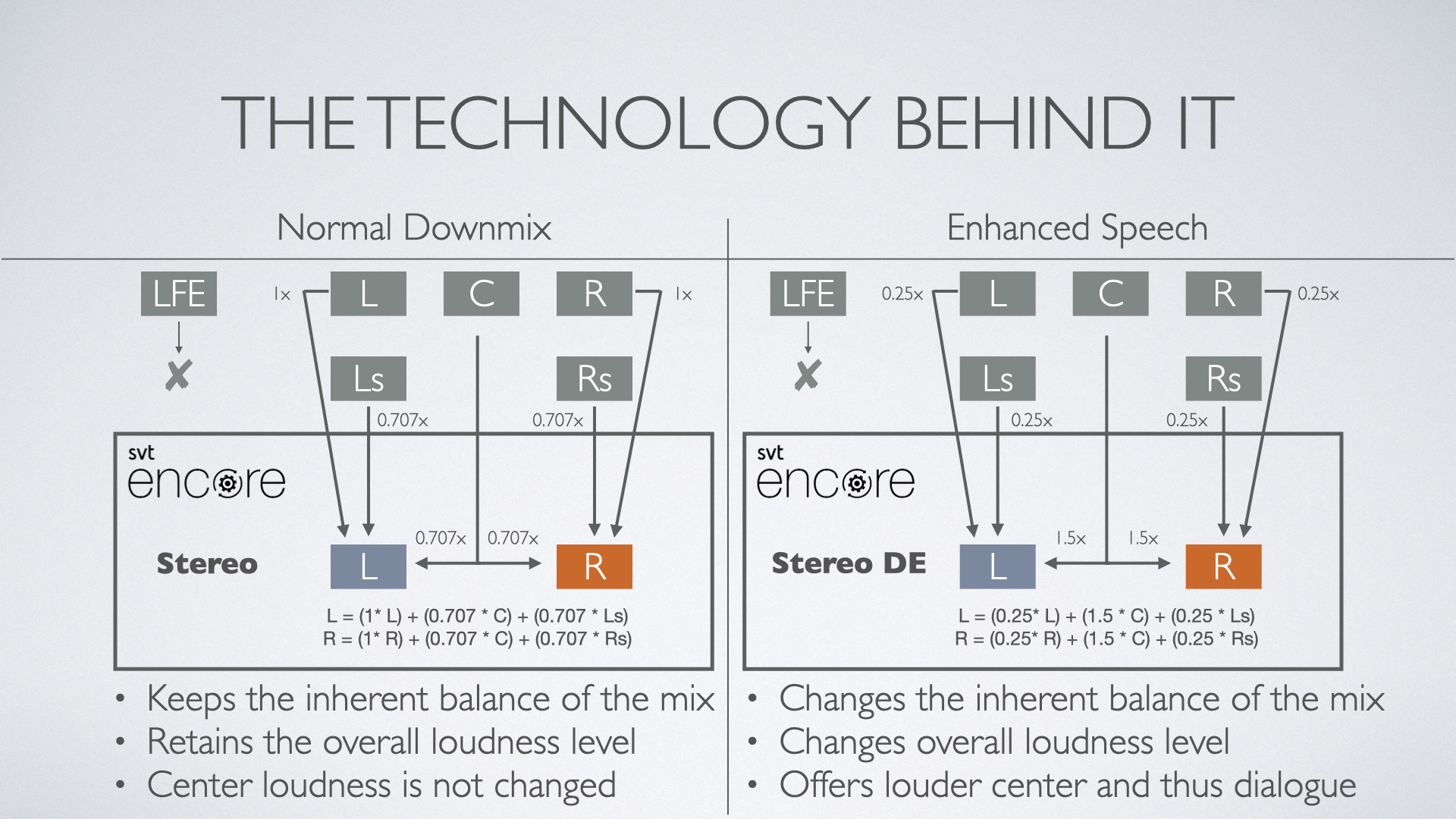}
\caption{Downmix Matrix} \label{downmixmatrix}
\end{figure}

\noindent When creating a 5.1 Enhanced Speech track the matrix is:
\[L_{ES} = 0.4* L\]
\[R_{ES} = 0.4* R\]
\[C_{ES} = 1.5* C\]
\[LFE_{ES} = 0.25 * LFE\]
\[Ls_{ES} = 0.25* Ls\]
\[Rs_{ES} = 0.25* Rs\]

\noindent Furthermore, the filter relies entirely on the assumption that the input audio streams have a standard channel layout, which is to say:

\noindent 3 channels : L, R, C\\
6 channels : L, R, C, LFE, Ls, Rs\\
8 channels : L, R, C, LFE, Ls, Rs, StereoLeft, StereoRight

Although the parameter values for the panning filter might seem intuitive, they are far from arbitrary, since they were devised through empirical testing. Initially we assumed that it would be sufficient to reduce the gain of every channel except the center, the idea being that by lowering the volume of everything else, the dialogue would stand out psychoacoustically. However, the effect turned out to be insufficient as the dialogue was still perceived as hard to hear. So, we created several variants of the same set of media clip, with different parameter values respectively, and made them available among our public test videos (https://www.svtplay.se/testbild). From there, we could easily test the results on several devices and in various viewing settings using a wide enough range of test material. Our findings made it clear that the optimal ratio for our purposes was achieved at 1.5x increase in gain of the center channel.

\section{Results}

\subsection{First programs}

In May 2020, SVT decided to take the experiment further and test it in the SVT Play VOD service. Three program titles was prepared with the ES feature, and published on svtplay.se: "Caravaggio, The Soul and the Blood", "Sommaren 85" and "Värsta listan". A survey link was put on these pages. The survey was also spread among members of The Swedish Association of Hard of Hearing People, Tillgänglighetstruppen and people who called or emailed SVT with audio complaints. The 2020 survey was published between May 13 and November 9 and got 86 responses in total. It included a brief explanation of the test procedure, and asked about age, sound in SVT programs, Enhanced Speech episode watched, opinion on and differences regarding the default audio and the ES audio in the episode, opinion about the ES feature and the likelihood of using the feature in the future.

The result was very positive, in all age groups. Enhanced Speech got the rating 8.3/10, compared with the default sound, which got 5,6/10. People said the difference between the sound was 8.2/10 and on the question whether they would use ES if it was available on every SVT program, they answered 9,0/10. As a comparison, Spoken Subtitles (Uppläst undertext in Swedish), was released in SVT Play in April 2021. A similar survey was done, with the main question: What is your opinion of the Spoken subtitles? The result was 5.8/10, with 21 participants. So, SVT's visitors gave ES a much higher rating compared to another new and anticipated accessibility feature. The survey participants commented for example: "Wonderful! Finally, I can watch TV without having to use subtitles.", "I don't need as high volume to hear what is said, so I don't disturb my neighbours.", "Excellent for me, I have severe tinnitus." or "An option to use when you have a bad speaker."

\subsection{Scaling up}

Because of the positive feedback from the audience, SVT continued to produce more programs with ES. The standard contracts for buying or ordering new shows was updated to include requirements on multi-channel sound. The internal sound mixing process was changed, to make sure that the center channel included only the voice. The coding process of the multi-audio programs was automated from the beginning of May 2021, so that SVT could offer many more titles each week. All shows with 5.1 or 3.0 sound now gets Enhanced Speech, if SVT doesn't manually exclude them. On June 21 2021, there were 128 shows with the ES feature available, out of around 2,300 in total. 

\subsubsection{Follow-up survey}
A follow-up survey was conducted between May 4 and June 21 2021, after the automated production of ES programs had started. This survey had 20 participants, and the result was similar to the first one, with a score of 8.4/10 score for the ES feature, and 8.7/10 on the probability of future usage.

\subsubsection{Effect on audio complaints}

As the number of titles with alternative audio has increased, the video starts with Enhanced speech have gone up by 431 percent, from 2,721 to 14,452 per week, and the complaints have gone down by 91 percent, from 104 to 9 per week, from April 12 to June 27.

\begin{figure}
\includegraphics[width=\textwidth]{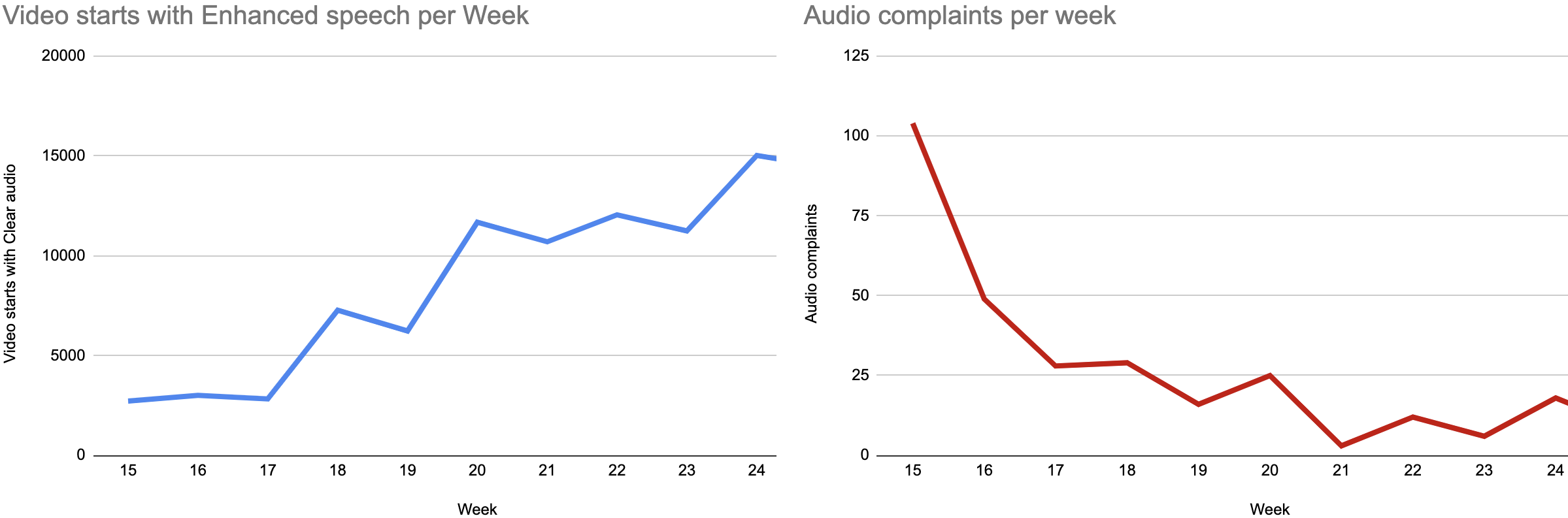}
\caption{Left: Video starts with Enhanced speech per week, May-June 2021. Right: Audio complaints per week, May-June 2021} \label{speechaudio}
\vspace{-3em}
\end{figure}

\subsection{Limitations}

The Enhanced Speech process assumes that all of the notional dialogue resides in the center channel, which wasn't always the case. The blurred center channel greatly reduced the benefit of the ES feature. But from May 2020 onwards, SVT reserved the center channel for notional speech in new programs. Another limitation is the static nature of the mix and lack of dynamic processing. Since all gain levels are adjusted as per the downmix matrix, the sections of a program that completely lack dialogue become very quiet. This effect gets stronger in some programs, depending on the balance of speech and background sound. By design, the Enhanced Speech mix warps the creative intent of an audio mix to a lesser or greater extent, and although the actual effect can still be a net benefit to the viewer, it is nonetheless important to emphasize this distortion. Moreover, the conversion process can only use multichannel audio as input, the procedure does not handle mono to mono or stereo to stereo. This means only multichannel audio programs got coded with Enhanced Speech. During 2020 and 2021, these programs were high-end and popular productions. This might have skewed the experience of these shows in a positive way, compared to others, with stereo sound.

ES is so far only available for VOD programs in SVT Play, not on broadcast, live streams or online channels, due to technical limitations. This means that some programs that get many audio complaints, like News and Sports, were not included. Also, SVT decided to avoid ES in programs based on music and singing, because it can give an unpleasant experience to hear the voice without musical instruments. Regarding the target group, the surveys were open to anyone who visited play.se, and there wasn't any background question regarding hearing, so it's impossible to tell if the participants were representative of the main target group, people with hearing loss.

\section{Conclusions and Future Work}
More research on Enhanced Speech in TV programs is needed. So far, SVT’s main target group has been the hard of hearing community, and the Enhanced Speech feature was tailored to be compatible with hearing aids, and used multichannel sound. This direction could be explored further, by adapting the mixing levels for different program genres, refining what sounds should be present in the center channel and letting users try out multiple levels and variants of static downmixing.

For people not using hearing aids, two alternative solution types could be explored, Dynamic Downmixing and Parametric Equalization (EQ). Dynamic downmixing means background noise is lowered when a recognizable voice is present. This technique is already used in the Spoken Subtitles feature at SVT. With Parametric EQ, frequencies present in the voice range are boosted. Both of these techniques involve the problem of recognizing what actually is a voice, and AI could be a big help here.

Even if sound complaints have decreased radically at SVT, some remain. They often concern News and Sports, and other programs with stereo sound. That means solutions for stereo sound, as those mentioned above, are important to consider for SVT. These solutions might not reach the same speech intelligibility as in SVT’s version of Enhanced Speech for multi-channel sound, but they could still bring a lot of value for viewers of the stereo sound content. Another area of exploration is how sound alternatives should be presented in the UI to be discovered by the users who need them. SVT is open for collaborations in all areas mentioned above, that may improve the users’ viewing experience.

The solutions explored in this paper could also be applied to adjacent industries, for example radio, podcasts and computer games. It is our hope that they will contribute to the creation of a more accessible and enjoyable soundscape in people's lives.

\clearpage
%
%
%

\bibliographystyle{splncs04}
\bibliography{bibliography}

\end{document}